\documentstyle[epsf,twocolumn,prl,aps]{revtex}
 \newcommand{\figwidth}{3.375in} 
\begin{document}
\draft

 \twocolumn[\hsize\textwidth\columnwidth\hsize\csname @twocolumnfalse\endcsname

\title{
Phase separation at all interaction strengths in the {\em t-J} model}
\author{ C. Stephen Hellberg\cite{steve} and E. Manousakis\cite{stratos} }
\address{
Department of Physics and Center for Materials Research and Technology, \\
	Florida State University, Tallahassee, FL 32306-3016 }
\date{\today}
\maketitle
\begin{abstract}
\noindent
We investigate the phase diagram of the two-dimensional \mbox{\em t-J} model
using a recently developed Green's Function Monte Carlo method
for lattice fermions.
We use
the technique to calculate exact ground-state energies
of the model on large lattices.
In contrast to many previous studies, we find the model
phase separates
for all values of $J/t$.
In particular, it is unstable at the hole dopings
and interaction strengths at which the model was thought
to describe the cuprate superconductors.
\end{abstract}
\pacs{PACS numbers:  71.10.+x, 71.45.Gm, 74.20.-z}

 ]

Models idealizing the environment of interacting constituents
are often used
to understand collective behavior.
In many cases the problem of providing a solution to
a model, even though in the model
one has abstracted the minimal complexity only, 
is overwhelmingly difficult.
For example, the antiferromagnetic Heisenberg model,
\begin{equation}
   H_{Heis}  =  J \sum_{\langle ij \rangle} {\bf S}_{i} \! \cdot \! {\bf S}_{j},
\label{heis}
\end{equation}
on an infinite square lattice,
where the $\bf S_i$ are spin-$\frac{1}{2}$ operators, 
was thought to describe the undoped parent compounds
of the superconducting
cuprates.
Despite the simplicity of the model, it has no exact solution.
However,
since the model has only spin degrees of freedom,
numerical solution to the model with satisfactory accuracy has become 
possible.
The physical picture emerging from the Heisenberg model
and
from
extensions obtained by adding small correction terms to it
(which can be understood perturbatively),
is in good agreement with 
the experimental results of the undoped
cuprates\cite{manousakis91}. 

The next step is to examine
the effect of introducing holes in a minimal
way into a Heisenberg antiferromagnet.  
The \mbox{\em t-J} model is perhaps the simplest abstraction to describe the
environment experienced by holes in the limit of strong
on-site Coulomb repulsion.
The \mbox{\em t-J} Hamiltonian, on a square lattice,
is written in the subspace with no doubly occupied sites as
\begin{equation}
   H
   = - t
      \sum_{\langle ij \rangle \sigma}
      ( c_{i\sigma}^{\dagger}c_{j\sigma}
		+ {\rm h.c.}
		)
 +
      J
\sum_{\langle ij \rangle}
      ( {\bf S}_{i} \! \cdot \! {\bf S}_{j} -
         \frac{n_i n_j}{4} ) ,
\label{tj-ham}
\end{equation}
where
$c_{i\sigma}^{\dagger}$ creates an electron of spin $\sigma$ on site
$i$, and $n_i = \sum_\sigma c_{i\sigma}^{\dagger}c_{i\sigma}$.
One can hope, by adding correction terms to such a model,
some day to understand the doped materials.
It is clear, however, that
no progress can be made without the capacity to understand the simple first.

Even though there are numerous previous studies\cite{dagotto} of the 
\mbox{\em t-J} model, its phase diagram is unclear.
More importantly, 
because of the inherent problem in drawing unique conclusions from small
size numerical calculations, the ambiguity and the confusion of the community 
due to such studies grows.
One of the main discrepancies concerns the phase-separation boundary of
the model.
Emery, Kivelson, and Lin (EKL)
\cite{emery90}
used a combination of analytic and numerical calculations
to argue for the existence of phase separation at all interaction strengths.
Subsequently
many other groups \cite{putikka,fehske91,poilblanc,gros92,kohno}
examined larger systems numerically and
found the model phase separates only for $J \gtrsim t$,
i.e., outside the physical region of the model,
and the results of EKL were strongly questioned.
Most of these studies used a vanishing inverse
compressibility as the criterion for the onset of phase
separation \cite{putikka,fehske91,poilblanc}.
The compressibility, however, is not the proper observable to
find the phase-separation boundary in the two-dimensional \mbox{\em t-J} model,
where the transition is first order.
It is true that the compressibility diverges in the region of phase separation, 
but it jumps discontinuously at the boundary with the uniform phase.
Numerically, this discontinuity is difficult to see in even
large finite systems due to
the surface energy of the two coexisting phases. 
The surface raises the energy of a phase-separated system,
and we find the inverse compressibility
remains positive
even where the system
phase separates.

In this paper we
calculate the phase-separation boundary of the \mbox{\em t-J} model using 
the Maxwell construction 
which
suffers very little from finite-size effects. 
We present results
on the ground states of significantly larger size systems
than could be studied previously by using a new powerful numerical technique.
We find a phase
diagram for the \mbox{\em t-J} model exhibiting phase separation at all
interaction strengths and in particular in the physical region of the
model.
This result confirms the conjecture of EKL and contradicts the more
accepted phase-separation boundary.
We propose a phase diagram of the model
consistent with all the available reliable results
and discuss the consequencies of this phase diagram for improved models of the
copper oxides.

For sufficiently large $J$, the \mbox{\em t-J} model phase separates completely
due to the attractive nature of the interaction term in (\ref{tj-ham}).
The two phases have electron densities per site of one and zero, respectively.
The energy per site in the electron region is 
$e_H \simeq -1.16934 J$ as determined by
calculations on the Heisenberg model \cite{manousakis91,runge92}.

As $J/t$ is reduced, $s$-wave
electron pairs evaporate from the high-electron density
region for $J < J_c \simeq  3.4367 t$\cite{emery90,hellberg95,hellberg96}.
Isolated larger clusters are never stabilized.
In this range, the two phase-separated regions contain
all electrons (no holes)
and some electrons (some holes).

To determine the phase-separation boundary for $J<J_c$, we use
a recently developed Green's Function Monte Carlo method to calculate
the ground-state energy of the model on large finite lattices
\cite{hellberg95,hellberg96,hellberg93,boninsegni92,boninsegni93,chen95}.
Prior to this work, the ground states of correlated fermions on large 
lattices had been calculated exactly only in one dimension \cite{hellberg93}
or in the limits of small numbers of holes \cite{boninsegni92,boninsegni93}
or electrons \cite{hellberg95}.
Starting from an initial trial state, this method projects the state
iteratively onto its lowest energy eigenstate.
In principle, this technique could be used to project any trial state that
overlaps the ground state, but trial states with larger overlaps yield
smaller statistical errors and require less computer time.
We use generalized singlet Jastrow-Slater and Jastrow-paired
wave functions as initial
trial states and have verified convergence to the ground
state by comparing with exact results on small systems
and by checking that different trial states converge to the
same state\cite{hellberg95}.
Trial states with exotic pairing states or broken time-reversal
symmetry were never needed to converge to the ground state.
The method suffers from the negative-sign problem,
which causes the statistical fluctuations to diverge exponentially with
increasing system size at a fixed density.
Fortunately,
with carefully chosen initial trial and
guiding functions, we are able to calculate the ground state energies
of significantly larger systems than
are possible with exact diagonalization.


To calculate the phase-separation boundary, we need the
behavior of the ground-state energy as a function of density.
For a given finite system size,
the shape of the Fermi surface changes dramatically as the electron number is
changed, resulting in seemingly random oscillations in the energy.
To avoid these shell effects, we vary the density
by changing the system size, keeping the electron number and
shape of the Fermi surface constant.

\begin{figure}[ht]
\epsfxsize=\figwidth\centerline{\epsffile{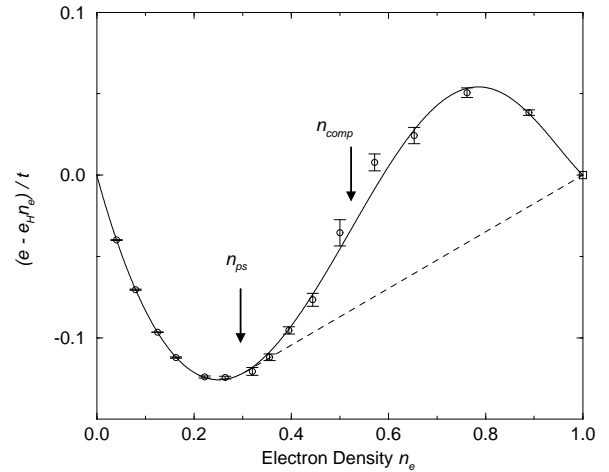}}
\caption{
The ground-state energy per site at $J = 2.5 t$ for $32$ electrons.
For clarity, the energies are shifted by a linear factor, $ - e_H n_e$.
The circles with error bars show the energies calculated on lattices
of dimensions $6 \times 6$, $7 \times 6$, ..., $28 \times 28$.
A sixth-order polynomial fit to the data is shown as the solid line,
which is extended
to
the Heisenberg energy, the square at energy zero in this shifted plot.
The dashed line shows the ground-state energy of the infinite system in the
phase-separated region.
We find the onset of phase separation occurs at
$n_{ps} = 0.296 \pm 0.004$,
while the inverse compressibility vanishes at $n_{comp} = 0.52 \pm 0.10$.
}
\label{fig:extrap}
\end{figure}

The ground-state energy
at $J=2.5t$ for 32
electrons on a variety of system sizes is shown in Fig.\ \ref{fig:extrap}.
These finite systems necessarily constrain the electron density to be uniform
on the length scales of the system size.
We fit the discrete data to a polynomial, $e(n_e)$, shown as the solid curve,
in order to treat the energy as a continuous function of density.
The dashed line, $e_{ps}(n_e)$, is a linear function that intersects
the Heisenberg energy, $e_H$ at electron density $n_e=1$ and intersects
$e(n_e)$ tangentially at a density labeled $n_{ps}$.

It is straight forward to show that the ground state of the infinite
system at a density $n_e>n_{ps}$ cannot be a uniform phase, because
the energy of the uniform phase, $e(n_e)$, is higher than $e_{ps}(n_e)$
at the same density.
This latter energy corresponds to the energy of a mixture of two phases,
one at electron density $n_A = 1$ and the other at electron density
$n_B = n_{ps}$.
Therefore the infinite system phase separates into two regions with densities
$n_A$ and $n_B$, and its ground-state energy is given by $e_{ps}(n_e)$,
the value of the dashed line at the average density of the system.
This is known as the Maxwell construction \cite{emery90}.

The energy of the infinite system is given by the solid line in
Fig.\ \ref{fig:extrap} for $n_e<n_{ps}$ and by the dashed line
for $n_e>n_{ps}$.
This Maxwell construction
differs from that commonly used since the density of one of the
constituent phases
lies at an extreme limit of the allowed density range.
It is not possible to add electrons to the Heisenberg solid,
which has one electron on every site,
so the dashed line is not tangent to the fitting curve
at $n_e = 1$.
If the {\em t-J} model did allow electron densities $n_e>1$,
then the intersection point of the solid and dashed lines
would be shifted to higher densities where the curves could intersect
tangentially.

In order to be stable, the energy of the infinite system must be concave
everywhere.
Given the solid line in Fig.\ \ref{fig:extrap} and the allowed density
range of the {\em t-J} model, the dashed line drawn in the figure is
the only line possible to make the energy of the infinite system
globally concave.



We never examined systems with densities $n_e \gtrsim 0.94$,
so we cannot exclude
the reentrance of a homogeneous phase in this region.
For such a phase to be stabilized,
the solid curve in Fig.\ \ref{fig:extrap} would have
to drop back below the
dashed line in this density range.
The new Maxwell line would lie slightly below the one drawn and would
be tangent to the solid curve at both intersections, but
the phase-separated region
would persist at densities $n_e \lesssim 0.94$.
We never saw any indication of this possibility at any interaction strength.

In the one-dimensional \mbox{\em t-J} model,
the compressibility diverges continuously at the transition
point in contrast to the discontinuous transition
in two dimensions\cite{putikka,ogata}.
We have verified that in one dimension,
the Maxwell construction yields the same 
phase-separation boundary as that calculated using the inverse compressibility.



\begin{figure}[ht]
\epsfxsize=\figwidth\centerline{\epsffile{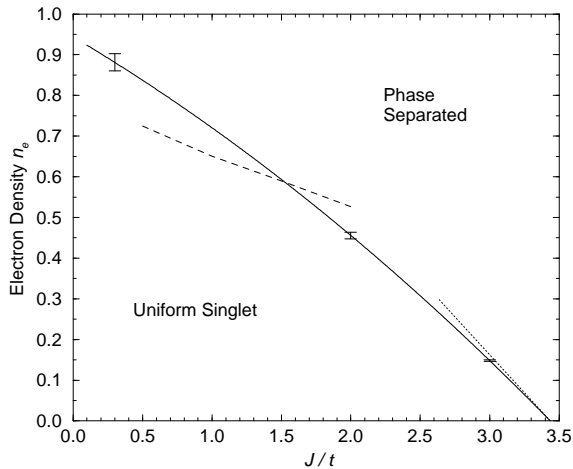}}
\caption{
The phase-separation boundary.
The solid line is a cubic fit to the phase boundary calculated
from systems with 32 to 60 electrons as described in the text.
Plotted are three sample errors, which increase with decreasing $J/t$.
The dashed line shows the boundary calculated using just 8 electrons.
The dotted line is a calculation of the limiting behavior
of the boundary at small electron density.
}
\label{fig:pd}
\end{figure}

Fig.\ \ref{fig:pd} shows the phase-separation boundary calculated 
at many interaction strengths.
Each point is calculated from a Maxwell construction using a fixed
number of electrons, either 60, 52, 50, 42, or 32,
on at least four different lattices
of size $L \times L$ or $L \times L\!+\!1$ where $7 \le L \le 28$.
At many interaction strengths,
we duplicated the calculation for different electron numbers,
and always found the discrepancy in the critical density
to be comparable to the statistical errors.
We have no reliable results for $J \lesssim 0.1t$, where the
phase-separation boundary extends beyond our maximum electron density
of $n_e \approx 0.94$, calculated using 60 electrons on
an $8 \times 8$ lattice.

At small electron density the phase-separation boundary is given by 
$J_c(n_e \rightarrow 0) = J_c(n_e = 0) + \frac{\pi J}{e_H} t n_e$,
which is plotted as the dotted line in Fig.\ \ref{fig:pd}.
This expression is obtained by assuming that the kinetic energy of paired
electrons varies with density while the binding energy is invariant.
The calculated boundary approaches this slope at low electron density.

The dashed line in Fig.\ \ref{fig:pd} shows the phase-separation boundary
calculated using the Maxwell construction with
8 electrons on $4 \times 3$, $4 \times 4$, and $5 \times 4$ lattices.
The energies of these small systems may be calculated
with Green's Function Monte Carlo or exact diagonalization.
The discrepancy between these results and those
using 32 or more electrons is apparent,
but both boundaries extend to very small $J/t$.

The numerical calculation of
EKL
varied the electron density of a 16-site lattice,
resulting in both finite-size and shell effects
absent in the present work\cite{emery90}.
However, they also found phase separation at all
$J/t$.

\begin{figure}[ht]
\epsfxsize=\figwidth\centerline{\epsffile{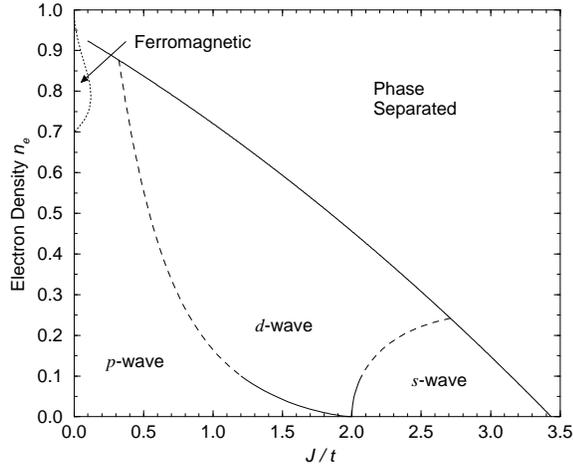}}
\caption{
Our proposed phase diagram.
The boundaries drawn as solid lines are accurate, while the others
are more speculative.
The phase-separation curve is the fit to the results 
shown in Fig.\ \protect\ref{fig:pd}.
The boundaries between phases of different
pairing symmetry
are calculated from expressions
in Ref.\ \protect\onlinecite{kagan94}.
They are accurate at low electron density, where they are drawn as solid
lines, and are extended as dashed lines
to the phase-separation boundary.
There is a ferromagnetic phase at very small $J/t$ and small hole dopings.
}
\label{fig:speculate}
\end{figure}

We now use our knowledge about the various regions of the 
phase diagram to derive a proposal for the full phase
diagram of the model, shown in in Fig.\ \ref{fig:speculate}.
We know with high degree of confidence
the phase-separation line obtained in this work, shown by the
solid line.
We also know what happens
(almost rigorously) at low electron 
density\cite{hellberg95,hellberg96,kagan94,chubukov93}.
In the limit of zero density, electrons form $s$-wave pairs
for $J > 2t$, and these pairs phase separate at  $J > J_c \simeq 3.4367 t$.
For infinitesimal densities,
the electrons are unstable to higher-angular-momentum pairings
due to the Kohn-Luttinger effect\cite{chubukov93}.
The strongest instability, as determined
by $T$-matrix calculations, is $p$-wave at small $J/t$
and $d_{x^2-y^2}$-wave at intermediate interaction
strengths\cite{hellberg96,kagan94}.
We have drawn the boundaries between these phases as solid lines
at small densities where the calculations are valid.
We believe these phases continue to higher 
densities but cannot trust the precise
location of the lines obtained by the low density expansion,
so we extend the boundaries as dashed lines.
We believe the lower boundary of the $d$-wave phase meets the
phase-separation boundary at $J^* \simeq 0.27 t$,
the minimum interaction strength for which a two-hole $d$-wave
bound state is stabilized\cite{boninsegni93}.
With the present work's evidence of phase separation,
we interpret the two hole binding
as indicating phase separation into a state of $d$-wave symmetry.
Therefore, we extend the lower boundary of the $d$-wave phase
using the expressions from Ref.\ \onlinecite{kagan94}
shifted linearly to extrapolate to $J^*$.

For $J \lesssim 0.1$ the model has a ferromagnetic instability,
which is sketched schematically as the dotted line
\cite{marder90}.
At low hole dopings,
this
boundary obeys $e_H = 2 \pi (1-n)^2 t$,
so $J_{ferro} \simeq 5.37 (1-n)^2 t$\cite{emery90}.
Series expansions indicate this phase does not extend beyond
$n_e \lesssim 0.7$ \cite{putikka92.2}.
In the unpolarized region,
a continuation of the low density $p$-wave pairing phase is compatible
with ferromagnetic correlations near the ferromagnetic instability.

This phase diagram will be sensitive to other terms that one may add to
the \mbox{\em t-J} Hamiltonian.
In particular, a long-range Coulomb repulsion will
suppress
macroscopic phase separation.
The local tendency for phase separation, however,
could have consequencies for the dynamics of the more complete Hamiltonian.
The competition of this tendency with the Coulomb forces
might create stable clusters with definite sizes,
as in nuclear droplets (nuclei),
or it might give rise to either the ``local''
collective modes, such those considered by Emery
and Kivelson \cite{emery93}, or to some superlattice structure,
such as those seen in neutron diffraction experiments \cite{tranquada97}.

In conclusion,
we have shown the two-dimensional \mbox{\em t-J} model phase separates
at some range of densities for $J \gtrsim 0.1t$,
and we believe the instability extends to all positive interaction strengths.
The only assumption made is
that the uniform phase is unpolarized in this region.
In particular,
phase separation occurs in the region of 
parameter space where the model is
thought to apply to the cuprate superconductors.
We verified that phase separation extends to electron
densities of at least $n_e \approx 0.94$ at all interaction strengths,
and we believe it extends to $n_e = 1$, the undoped antiferromagnet.
The main reason for the discrepancy of these results with previous work is that
the phase-separation boundary is
determined far more accurately using the Maxwell construction 
than the inverse compressibility.
Finally, we have proposed a complete phase diagram of the model
including pairing symmetries based on all accurate calculations
presently known to us.



We thank
N. E. Bonesteel,
Y. C. Chen,
T. K. Lee,
P. Monthoux,
W. O. Putikka,
A. W. Sandvik,
P. Schlottmann,
and
C. T. Shih
for useful conversations.
This work was supported by the Office of Naval Research Grant No.
N00014-93-1-0189. 


\vspace {-.1in}
\begin{references}
\vspace {-.6in}

\bibitem[\dag]
{steve} Electronic address: hellberg@dave.nrl.navy.mil\\
Present address: Complex Systems Theory Branch, Naval Research Laboratory,
Washington DC 20375
\bibitem[\ddag]
{stratos} Electronic address: stratos@ithaca.martech.fsu.edu



\bibitem{manousakis91}
E. Manousakis, Rev.\ Mod.\ Phys. {\bf 63},  1  (1991).

\bibitem{dagotto}
E. Dagotto, Rev.\ Mod.\ Phys. {\bf 66},  763  (1994);
E. Dagotto, and J. Riera, Phys.\ Rev.\ Lett.\ {\bf 70}, 682 (1993);
E. Dagotto, {\em et.\ al.}, Phys.\ Rev.\ {\bf B 49}, 3548 (1994).

\bibitem{emery90}
V.J. Emery, S.A. Kivelson, and H.Q. Lin, Phys.\ Rev.\ Lett. {\bf 64},  475
  (1990);
S.A. Kivelson, V.J. Emery, and H.Q. Lin, Phys.\ Rev.\ B
{\bf 42}, 6523 (1990);
S.A. Kivelson and V.J. Emery, 
in {\em Strongly correlated electronic materials:
the Los Alamos symposium, 1993},
edited by K.S. Bedell {\em et al.}\ (Addison-Wesley, Reading, CA, 1994).

\bibitem{putikka}
M.U. Luchini,
{\em et al.},
Physica C {\bf 185-189},  141  (1991);
W.O. Putikka, 
M.U. Luchini, and T.M. Rice, Phys.\ Rev.\ Lett. {\bf 68},  538  (1992).

\bibitem{fehske91}
H. Fehske, V. Waas, H. R\"oder, and H. B\"uttner, Phys.\ Rev.\ B {\bf 44},
 8473  (1991).

\bibitem{poilblanc} D. Poilblanc, Phys.\ Rev.\ B {\bf 52},  9201  (1995).

\bibitem{gros92}
R. Valenti and C. Gros,  Phys.\ Rev.\ Lett.\ {\bf 68}, 2402 (1992);
H. Yokoyama and M. Ogata, J.\ Phys.\ Soc.\ Japan {\bf 65}, 3615 (1996).

\bibitem{kohno} M. Kohno, Phys.\ Rev.\ B {\bf 55},  1435  (1997).

\bibitem{runge92}
K.J. Runge, Phys.\ Rev.\ B {\bf 45}, 12292 (1992);
N. Trivedi and D.M. Ceperley, {\em ibid.}\ {\bf 40},  2737  (1989);
J. Carlson, {\em ibid.}\ {\bf 40},  846  (1989).

\bibitem{hellberg95}
C.S. Hellberg and E. Manousakis, Phys.\ Rev.\ B {\bf 52},  4639  (1995).

\bibitem{hellberg96}
C.S. Hellberg and E. Manousakis,  in {\em Physical Phenomena at High Magnetic
  Fields - II},
edited by Z. Fisk {\em et al.}\ (World Scientific, Singapore, 1996).

\bibitem{hellberg93}
C.S. Hellberg and E.J. Mele, Phys.\ Rev.\ B {\bf 48},  646  (1993);
Y.C. Chen and T.K. Lee,  Phys.\ Rev.\ B {\bf 47},  11548  (1993).

\bibitem{boninsegni92}
M. Boninsegni and E. Manousakis, Phys.\ Rev.\ B {\bf 46},  560  (1992).

\bibitem{boninsegni93}
M. Boninsegni and E. Manousakis, Phys.\ Rev.\ B {\bf 47},  11897  (1993).

\bibitem{chen95}
Y.C. Chen and T.K. Lee,  Phys.\ Rev.\ B {\bf 51},  6723  (1995).


\bibitem{ogata}
M. Ogata, M.U. Luchini, S. Sorella, and F.F. Assaad, Phys.\ Rev.\ Lett. {\bf
  66},  2388  (1991).

\bibitem{kagan94}
M.Y. Kagan and T.M. Rice, J. Phys.\ Cond.\ Mat. {\bf 6},  3771  (1994).

\bibitem{chubukov93}
A.V. Chubukov and M.Y. Kagan, in {\em Physical Phenomena 
at High Magnetic Fields}, edited by E. Manousakis
{\em et al.}\ (Addison-Wesley, Redwood City, CA, 1992), p.\ 239;
A.V. Chubukov, Phys.\ Rev.\ B {\bf 48},  1097  (1993).

\bibitem{marder90}
L.B. Ioffe and A.I. Larkin, Phys.\ Rev.\ B {\bf 37},  5730 (1988);
M. Marder, N. Papanicolaou, and G.C. Psaltakis, Phys.\ Rev.\ B {\bf 41},  6920
   (1990);

\bibitem{putikka92.2}
W.O. Putikka, M.U. Luchini, and M. Ogata, Phys.\ Rev.\ Lett. {\bf 69},  2288
  (1992).


\bibitem{emery93}
V.J. Emery and S.A. Kivelson, Physica C {\bf 209},  597  (1993).

\bibitem{tranquada97}
J.M. Tranquada, J.D. Axe, N. Ichikawa, A.R. Moodenbaugh, Y. Nakamura,
and S. Uchida, Phys.\ Rev.\ Lett. {\bf 78}, 338 (1997).

\end{references}
\end{document}